\documentstyle[preprint,aps]{revtex}
\begin{document}
\draft
\bibliographystyle{prsty}

\title{\bf The BFV Approach for a Nonlocal Symmetry of QED }
\author{Silvio J. Rabello\thanks{e-mail: rabello@if.ufrj.br}\,
and Patricio Gaete\thanks{Present adress: Instituto de F{\'\i}sica,
Universidade Estadual do
Rio de Janeiro, Rua S\~ao Francisco Xavier 524, Rio de Janeiro  RJ,
CEP 20550-013 Brasil}}

\address{\it Instituto de F\'\i sica, Universidade Federal do Rio de
Janeiro, Rio de Janeiro,  RJ \\ Caixa Postal 68.528-CEP 21945-970, Brasil}

\date{\today}

\maketitle

\begin{abstract}

{\sl In this paper we use the Batalin-Fradkin-Vilkovisky formalism to study
a recently proposed nonlocal symmetry of QED. In the BFV extended phase
space we show that this symmetry  stems from a canonical transformation in
the ghost sector.}

\end{abstract}

It is well known that gauge symmetry plays an essential role in the modern
picture of quantum field theory, where the locality of the fields is implicit.
Thus, in order to carry out a consistent quantization of a theory with gauge
symmetry, it is necessary to eliminate the non-physical degrees of freedom.
The usual procedure to achieve this is to impose auxiliary conditions
(gauge conditions) and their form is dictated largely by computational
convenience.

In the path integral quantization methods (both Lagrangian and Hamiltonian)
the gauge invariance is included by means of the extension of the phase
space incorporating the Grassmann odd ghost variables. In this case ,
the basic idea is to replace the local gauge symmetry by a global
Becchi-Rouet-Stora-Tyutin supersymmetry (BRST)\cite{BRST}. Today this
BRST symmetry plays a crucial  role in the quantization of gauge theories.

In this connection in recent times a great deal of attention has been devoted
to the study of new symmetries of QED.  Recently, Lavelle and McMullan
\cite{LaMc} found that QED exhibits a new nonlocal and noncovariant graded
symmetry, even so nilpotent. In such a case the symmetry  transformations
are compatible with the gauge fixing conditions. Moreover these authors
claim that this new symmetry may be used to refine the characterization of
the physical states given by the BRST charge.  Following these lines, Tang
and Finkelstein \cite{TaFi} constructed a covariant graded symmetry for QED
of which the noncovariant symmetry of Lavelle and McMullan is a special case.
Also Yang and Lee \cite{YL} derived a noncovariant but local symmetry of QED.
The Noether charges for all these symmetries are nilpotent and impose
constraints on the physical states. Nevertheless, these new symmetries
reported up till now are all constructed from a Lagrangian point of view.

On the other hand, the Hamiltonian formalism developed by Batalin, Fradkin
and Vilkovisky (BFV) \cite{BV} provides a powerful method for the BRST
quantization of constrained systems, e.g., QED.  As it is well known,
some properties of the BFV approach are: it does not need an auxiliary field
and the BRST transformations are independent of the gauge conditions,
it uses an extended phase space in which  the Lagrange multipliers and
ghosts are introduced as dynamical variables. The BRST charge for  first
class constrained systems can be constructed directly from the constraints.
Thus, the  algebraic structure of the constraints is captured in the BRST
charge and its nilpotency in a gauge independent way, unlike the Lagrangian
formulation in which the BRST charge is calculated from a gauge fixed
Lagrangian via Noether's theorem.

It is therefore motivating in this context to consider the question of
how arises, in the Hamiltonian approach, the Lavelle and McMullan's
symmetry. In order words we want to find the relation between the Lagrangian
and Hamiltonian symmetry generators.  A detailed discussion of the BFV
formalism can be found in \cite{HeTe}. We here collect some results of the
BFV treatment for QED.  The Lagrangian for QED with photon gauge field $A_\mu$
and Dirac electron field $\psi$ is given by \footnote{$\hbar=c=1$ and
$\eta_{\mu\nu}=diag(1,-1,-1,-1)$}
\begin{equation}
\label{lqed}
{\cal L}=-{1\over 4}F_{\mu\nu}F^{\mu\nu}+{\bar\psi}(i\gamma^\mu D_\mu -m)
\psi\,,
\end{equation}
where $F_{\mu\nu}=\partial_\mu A_\nu -\partial_\nu A_\mu$
and $D_\mu=\partial_\mu +ieA_\mu$. The canonical momenta of the gauge
fields are $\pi_\mu=F_{0\mu}$
with the only nonvanishing canonical Poisson brackets being:
\begin{eqnarray}
\label{cpb1}
[A^\mu({\bf x}, t),\pi_\nu({\bf y},t)]&=&\delta^\mu_\nu\delta({\bf x}-{\bf y})
\,, \\
\label{cpb2}
[ \psi ({\bf x},t),\psi^\dagger ({\bf y},t)]&=&-i\delta({\bf x}-{\bf y})
\end{eqnarray}
As we can see there is one primary constraint, $\pi_0=0$, and the canonical
Hamiltonian is
\begin{equation}
\label{HC}
H=H_0+\int d^3 x A_0(\nabla\cdot\mbox{\boldmath $\pi$}+e\psi^\dagger\psi)\,,
\end{equation}
where
\begin{equation}
\label{H0}
H_0=\int d^3 x \biggl(\psi^\dagger(i\mbox{\boldmath $\alpha$}\cdot{\bf D}
+\gamma_0 m)\psi +{1\over 2}\mbox{\boldmath $\pi$}^2
+{1\over 4}F^{ij}F_{ij}\biggr)\,.
\end{equation}
The conservation in time of the constraint $\pi_0=0$ gives us the Gauss law:
\begin{equation}
\label{G}
\nabla\cdot\mbox{\boldmath $\pi$}+e\psi^\dagger\psi=0
\end{equation}
There are no more constraints in the theory and the two we have found are
first class (they have a vanishing Poisson bracket). Having identified the
first class constraints we are ready to apply the BFV method where one
starts with the Gauss law constraint in the phase space $(A_i,\pi_i)$  and
introduces  the pair of anticommuting ghosts  $(C(x),{\cal P}(x))$ of ghost
number 1 and $-$1 respectively. Then one adds the Lagrange multiplier
$A_0(x)$ and its vanishing momentum $\pi_0$, to take care of this new
constraint we have the antighost pair $({\bar C}(x),{\bar{\cal P}}(x))$ of
ghost number $-$1 and 1 respectively. The Poisson algebra of these ghosts
is (the nonzero part of it)
\begin{equation}
\label{ghost1}
[C({\bf x},t),{\cal P}({\bf y},t)]=
[{\bar C}({\bf x},t),{\bar{\cal P}}({\bf y},t)]=-\delta({\bf x}-{\bf y})\,.
\end{equation}
In this extended phase space the generator of the BRST symmetry is given by

\begin{equation}
\label{BRSTC}
Q=-\int d^3 x \biggl[C\biggl(\nabla\cdot\mbox{\boldmath $\pi$}
+e\psi^\dagger\psi\biggr)+i{\bar{\cal P}}\pi_0\biggr]\,.
\end{equation}
The striking property of $Q$ is that $[Q,Q]=0$ off-shell, with this at
hand we can define a classical BRST cohomology for any gauge theory.
It follows from $[Q,Q]=0$ and the Jacobi identity for the Poisson
brackets that $[[{\cal F},Q],Q]=0$ for any ${\cal F}$.
{}From that we conclude that any BRST invariant quantity is defined modulo
a Poisson bracket $[{\cal F},Q]$.  In our case we have that $[H_0,Q]=0$
and so it is for the extended $H$:
\begin{equation}
\label{Hext}
H=H_0+[Q,\Psi]
\end{equation}
Here $\Psi$ is an arbitrary fermionic functional of the extended phase space
variables, that is chosen  to give dynamics to all fields and
to select a gauge slice. The extended phase space action is then
\begin{equation}
\label{sBFV1}
S_{BFV}=\int d^4 x\biggl({\dot A}_\mu\pi^\mu+i\psi^\dagger{\dot\psi}
+{\dot C}{\cal P}+{\dot{\bar C}}{\bar{\cal P}}\biggr)-
\int dt\biggl(H_0+[Q,\Psi]\biggr)\,.
\end{equation}
By construction this action is invariant under $Q$, that is,
under the global transformations
\begin{eqnarray}
\label{dBRST}
\delta A_i&=&\partial_i C\qquad,\qquad\delta A_0=i{\bar{\cal P}}\,,\\
\delta \pi_\mu&=&0\qquad ,\qquad\quad
\delta\psi=-ie C\psi\,,\\
\delta C&=&0\qquad ,\qquad\quad
\delta {\bar C}=i\pi_0\,,\\
\delta {\cal P}&=&\nabla\cdot\mbox{\boldmath $\pi$}+e\psi^\dagger
\psi\qquad ,\qquad
\delta {\bar{\cal P}}=0\,.
\end{eqnarray}

The Fradkin-Vilkovisky theorem \cite{BV} states that the path integral
\begin{equation}
\label{Z}
Z_{BFV}=\int [dA_\mu]\,[d\pi_\mu]\,[dC]\,[d{\bar C}]\,[d{\cal P}]\,
[d{\bar{\cal P}}]\,e^{iS_{BFV}}\,,
\end{equation}
has no dependence on $\Psi$. For QED a common choice of $\Psi$ is
\begin{equation}
\label{gfix}
\Psi=\int d^3 x \biggl[{\cal P}A_0-i{\bar C}(\nabla\cdot{\bf A}
-{\xi\over 2}\pi_0)\biggr]\,,
\end{equation}
where $\xi$ is a real parameter that describes a set of gauges,
e.g. for $\xi=0$ we have the Landau gauge, for  $\xi=1$ the Feynman
gauge, and for $\xi=\infty$ the unitary gauge. That choice of $\Psi$
give us the action
\begin{eqnarray}
\label{sBFV2}
S_{BFV}&=&\int d^4 x\biggl [-{\dot {\bf A}}\cdot\mbox{\boldmath $\pi$}
+i\psi^\dagger{\dot\psi}+{\dot C}{\cal P}+{\dot{\bar C}}{\bar{\cal P}}
\nonumber\\
&-&\biggl(\psi^\dagger(i\mbox{\boldmath $\alpha$}\cdot{\bf D}
+\gamma_0 m)\psi +{1\over 2}\mbox{\boldmath $\pi$}^2
+{1\over 4}F^{ij}F_{ij}\biggr)\nonumber\\
&-&A_0(\nabla\cdot\mbox{\boldmath $\pi$}+e\psi^\dagger\psi)
+{\xi\over 2}\pi_0^2+\pi_0 \partial_\mu A^\mu+i{\cal P}{\bar{\cal P}}
-i{\bar C}\nabla^2 C\biggr]\,.
\end{eqnarray}
Performing the functional integrals over the momenta we recover
the usual Faddeev-Popov action
\begin{equation}
\label{sFP}
S_{FP}=\int d^4 x\biggl[-{1\over 4}F_{\mu\nu}F^{\mu\nu}
+{\bar\psi}(i\gamma^\mu D_\mu -m)\psi
-{1\over 2\xi}(\partial_\mu A^\mu)^2
+i{\bar C}\partial_\mu\partial^\mu C\biggr] \,,
\end{equation}
with its  invariance under the BRST transformations
\begin{eqnarray}
\label{dBRST2}
\delta A_\mu&=&\partial_\mu C\qquad ,\qquad
\delta\psi=-ie C\psi\,,\\
\delta C&=&0\qquad ,\qquad
\delta {\bar C}=-{i\over\xi}\partial_\mu A^\mu \,.
\end{eqnarray}

Recently Lavelle and McMullan \cite{LaMc} discovered that the
Faddeev-Popov action is also invariant under the nonlocal variations:
\begin{eqnarray}
\label{dLaMc}
\delta^\perp A_i&=&i{\partial_i\partial_0\over\nabla^2} {\bar C}\qquad ,
\qquad
\delta^\perp A_0 =i{\bar C}\,,\\
\delta^\perp\psi &=&e\biggl({\partial_0\over\nabla^2}{\bar C}\biggr)
\psi\qquad ,\qquad
\delta^\perp C =A_0-{\partial_i\partial_0\over\nabla^2}A_i
+{e\over\nabla^2}
\psi^\dagger\psi\,,\\
\delta^\perp {\bar C}&=& 0\,.
\end{eqnarray}
This was found in the search for a symmetry that decreases the ghost number
by one and  that obeys $\delta^\perp (\partial_\mu A^\mu)=0$,
i.e., leaves the gauge fixing condition  invariant. And also
with  $[Q,Q^\perp]\neq 0$. This symmetry was discovered in the configuration
space. The question we pose is: How does it arise in the BFV phase space?
In the phase space formulation we have the freedom of performing
canonical transformations, in such a way that any two
BRST generators are related by  such transformations \cite{HeTe}.
We propose now the following canonical transformation in the ghost sector:
\begin{eqnarray}
\label{ct}
C'&=&{1\over\nabla^2}{\cal P}\qquad ,\qquad {\cal P}'=\nabla^2 C\,,\\
{\bar C}'&=&{\bar{\cal P}}\qquad , \qquad {\bar{\cal P}}'={\bar C}\,
\end{eqnarray}
It is easy to see that the BFV Lagrangian is form invariant under
these replacements, so that dropping the primes we have the new BRST
charge, that we call $Q^\perp$, being
\begin{equation}
\label{LaMcC}
Q^\perp=-\int d^3 x\biggl[
{1\over\nabla^2}{\cal P}\biggl(\nabla\cdot\mbox{\boldmath $\pi$}
+e\psi^\dagger\psi\biggr)+i{\bar C}\pi_0\biggr]\,.
\end{equation}
The action of this charge on the extended phase space is
\begin{eqnarray}
\label{dBRST3}
\delta^\perp A_i&=&{\partial_i\over\nabla^2}{\cal P}
\qquad,\qquad\delta^\perp A_0=i{\bar C}\,,\\
\delta^\perp \pi_\mu&=&0 \qquad ,\qquad\quad
\delta^\perp\psi=-ie {1\over\nabla^2}{\cal P}\psi\,,\\
\delta^\perp C&=&{1\over \nabla^2}(\nabla\cdot\mbox{\boldmath $\pi$}
+e\psi^\dagger\psi)\qquad ,\qquad
\delta^\perp {\bar C}=0\,,\\
\delta^\perp {\cal P}&=&0 \qquad ,\qquad\quad
\delta^\perp{\bar{\cal P}}=i\pi_0\,.
\end{eqnarray}
It is straightforward to see that on integration over the momenta
these transformations reduce to the ones found in \cite{LaMc}.
Among the huge number of possible canonical transformations, this
one meets the requirements of changing the ghost number of $Q$ to minus
one and that after integration over the momenta gives
$\delta^\perp (\partial_\mu A^\mu)=0$ off-shell.
To understand this last property recall that in the extended phase space
we have $\delta (\partial_\mu A^\mu)=i{\dot{\bar{\cal P}}}-\nabla^2 C $
which in configuration space translates into
$\delta (\partial_\mu A^\mu)=\partial_\mu\partial^\mu C$, but this is
just the classical equation of motion of $C$.
What the above proposed canonical transformation  performs is a swap
of the ghost Hamilton equations of motion, that is,
$\delta^\perp (\partial_\mu A^\mu)=i{\dot{\bar C}}-{\cal P}$,
which after integration over the momenta gives zero, turning the
variation of the gauge fixing condition from  null on-shell to null off-shell .

\vskip 0.5 cm
{\bf Acknowledgments}
\vskip 0.5 cm
The authors are grateful to C. A. Linhares for reading the manuscript.
P. Gaete thanks the Instituto de F\'\i sica of UFRJ for the hospitality.
This work was partially supported by CNPq (Brazilian Research Council).

\end{document}